# Possibility of Inducing $^{241}$Am Fluctuations

(Primakoff Photon-Magnetic Field Coupling Experiment)

C. Scarlett[a], E. Fischbach[b], B. Freeman[b], J. J. Coy[c], P. Edwards[d], D. Osborne[a], J. Edwards[a], L. Mwibanda[b], A. Alsayegh[b], and M. Chiu[e]

[a] Department of Physics, Florida A&M University

[b] Department of Physics and Astronomy, Purdue University

[c] Department of Computer Science, Ball State University

[d] Department of Physics, College of Coastal Georgia

[e] Brookhaven National Laboratory (BNL)

## ABSTRACT:

This paper reports on results from an experiment designed to search for exotic particles interacting with nuclear matter. These particles could be created through the Primakoff coupling between photons and an external magnetic field. Theory suggests this coupling leads to the production of weakly interacting particles (e.g. axions) that are important to understanding the lack of a measured neutron electric dipole moment (nEDM). The current experiment has been run to look for evidence of weakly interacting particles, created by photons propagating through a magnetic field, by studying their influence on the measured decay spectrum of Americium ($^{241}$Am). The results shown here reflect a statistically significant difference (sigma > 6.0) between observed decays when the experiment was run in a mode that allowed photons to traverse a magnetic field (light mode or sP for system-Photons) when compared to a second mode where the light was blocked from entering the cavity (dark mode or sD for system-Dark). This difference was observed to impact the count rate for the release of a 59.54keV gamma from $^{237}$Np. Repeated experimentation suggests the effect is robust and not due to spurious changes in background events. This could be confirmation that the Primakoff mechanism has been observed for visible photons. As importantly, this experiment looks at the possibility to develop a novel nuclear instrument that can modify nuclear decay rates.

## Introduction:

The absence of strong CP violations, given that such effects have been measured for weak interactions, has long puzzled particle physicists. Adding to the mystery is the apparent lack of a measurable neutron electric dipole moment (nEDM), which should arise from the quark content of the nucleons. An eloquent solution to both of these observations was proposed by Peccei-Quinn[1] as the coupling between a scalar field and at least one fermionic particle. This scalar field also gives rise to a boson particle dubbed the "axion." A number of experiments have been proposed and executed to search for this new particle, which in theory should result from interactions between photons and external magnetic fields. The photon-magnetic field interaction that could directly produce axions had already been theorized by Henry Primakoff as far back as 1951 to explain photo-production of mesons in the Coulomb field of a nucleus[2].

The presence of a magnetic field in space breaks the isotropy of space. Combined with the Peccei-Quinn scalar field, this broken isotropy should lead to absorption and/or polarization rotation whenever a photon beam passing through a magnetic field. Experimental searches for axion production and decay have largely focused on finding the lowest order Feynman diagram whereby photons beams rotate or disappear leading to optical effects that could be measured (see Figure 1 Left). After numerous experimental searches (PVLAS, BNL E840, BMV, etc.)[3-5] no group has yet seen an effect that demonstrates the production of axions.

The experiment described here takes a look at a higher order effect (see Figure 1 Right) where the production of axions and/or weakly interacting particles is just part of the equation. The current approach aims those weakly interacting particles at a radioactive sample to search for changes in the detected number of decays. Such a process involves two Feynman diagrams, one for the production of and the other for the absorption of these particles by a nucleus.

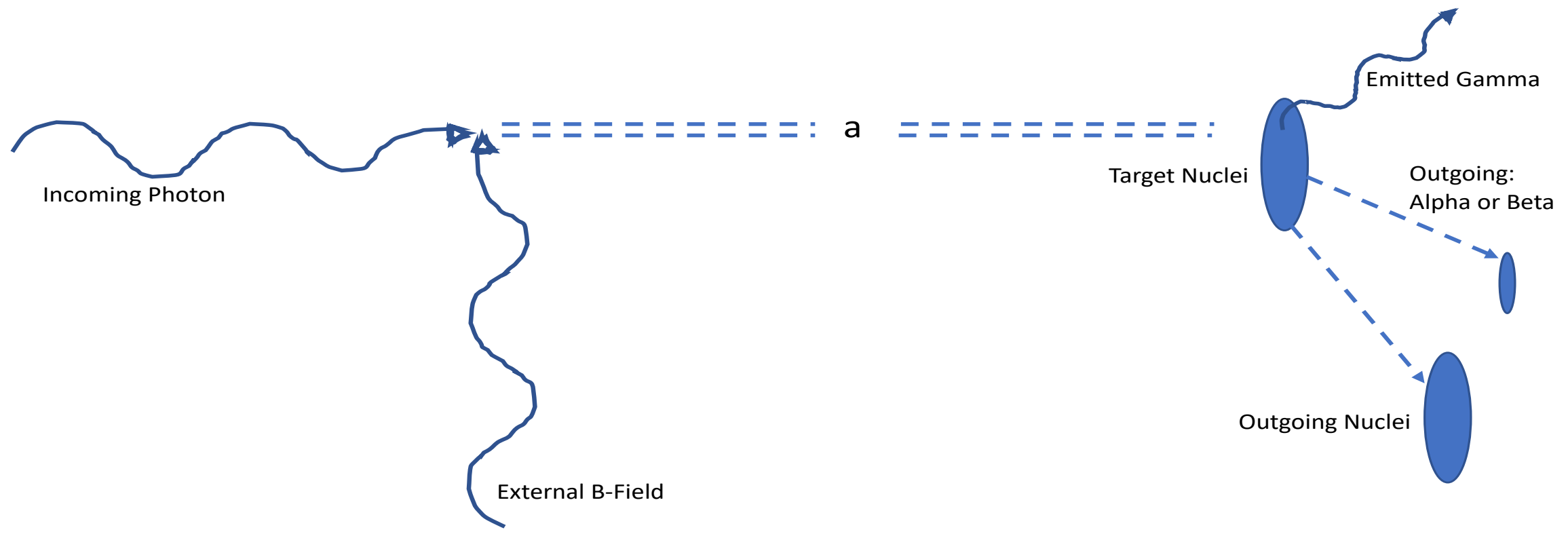

Figure 1: Primakoff production of axions (Left) and axion coupling to nuclear material (Right).

## $^{241}$Am:

$^{241}$Am results from a beta decay of plutonium ($^{241}$Pu). The isotope has a half-life of about 432 years. Notably it is one of the few isotopes that finds use in industry, in homes across the country to detect smoke. $^{241}$Am undergoes alpha decay to produce $^{237}$Np in an excited state. The excited $^{237}$Np emits intense gamma lines with energies of 59.54 and 26.32 keV. These lines are often used to calibrate nuclear detectors such as the NaI detectors used in this experiment.

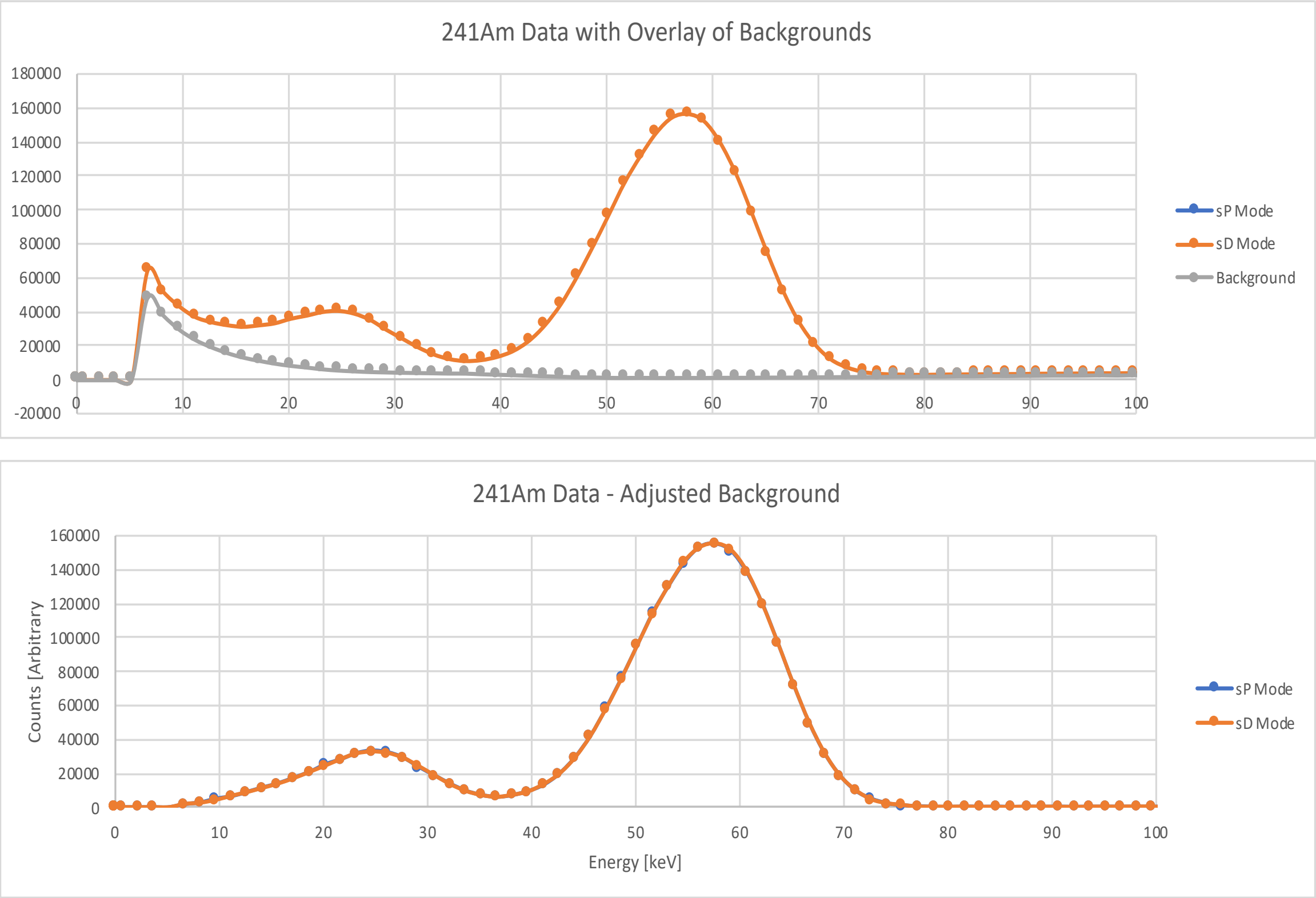

Figure 2: $^{241}$Am decay energy spectrum showing before (top panel) and after (bottom panel) background (top panel in gray) subtraction, collected using NaI gamma detectors.

Figure 2 shows a sample of data from the detection of $^{241}$Am decays. The plots references “sP” and “sD” modes to distinguish the experimental conditions described more below. At the resolution of this plot, there is no detectable difference between these modes. The analysis shown below reveals differences at a scale of ~ 0.33% or ~ 10 thousand out of 3 Million.

## Experimental Setup:

To detect any weakly interacting particles produced via the Primakoff mechanism, we propagate a beam of photons through a cavity which uses neodymium magnets to create a strong magnetic field. At the exit of the magnetic cavity, barriers prevent the light from directly interacting with the any of the radionuclides placed upstream where any weakly interacting particles should flow if produced in this process. The data is subdivided into an optical mode where the beam is allowed to pass through the cavity ("sP" indicated throughout this paper) and a dark mode where the beam is blocked before entering the cavity ("sD" indicating throughout this paper). The laser beams are always on to prevent any electronic effects associated with changing power drawn by turning on-off the beam.

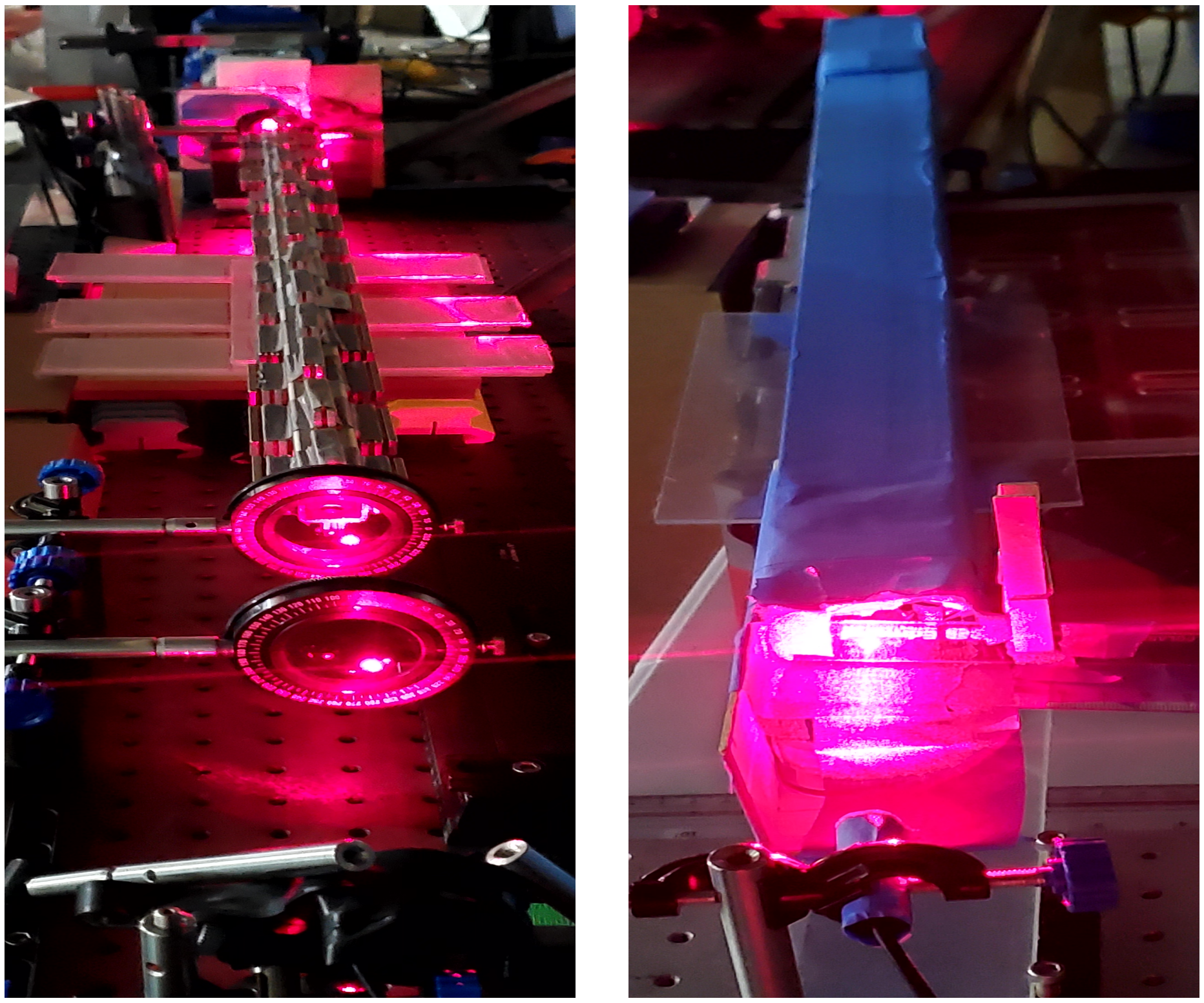

Figure 3: Magnetic cavity showing beams propagating through the magnetic fields.

Figure 3 shows two of the magnetic field apparatus used in taking the data analyzed in this work. The data is taken in an optical mode (sP) for 1 hour followed by a dark mode (sD) for 1 hour and then oscillated thereafter. A second detector is used to monitor the background radiation levels throughout the process. Because the second detector has a different, natural calibration relative to the detector used for the experimental data collection, the experimental detector is run in a "No Source" mode to collect both sD and sP data for background subtraction. In this way, the second detector is only used to establish the relative background levels, that are then used to set the amplitude for the No Source data.  It is the No Source data taken on the experimental detector that is subtracted to yield the $^{241}$Am energy spectrum to calculate the final results.

The magnetic field is position to be either aligned in the plane of the optical bench and perpendicular to the propagation direction of the beam, this is "BH" or B-field horizontal or the field is aligned perpendicular (vertical) to both the plane of the optical bench and the direction of propagation for the optical beam "BV" mode.

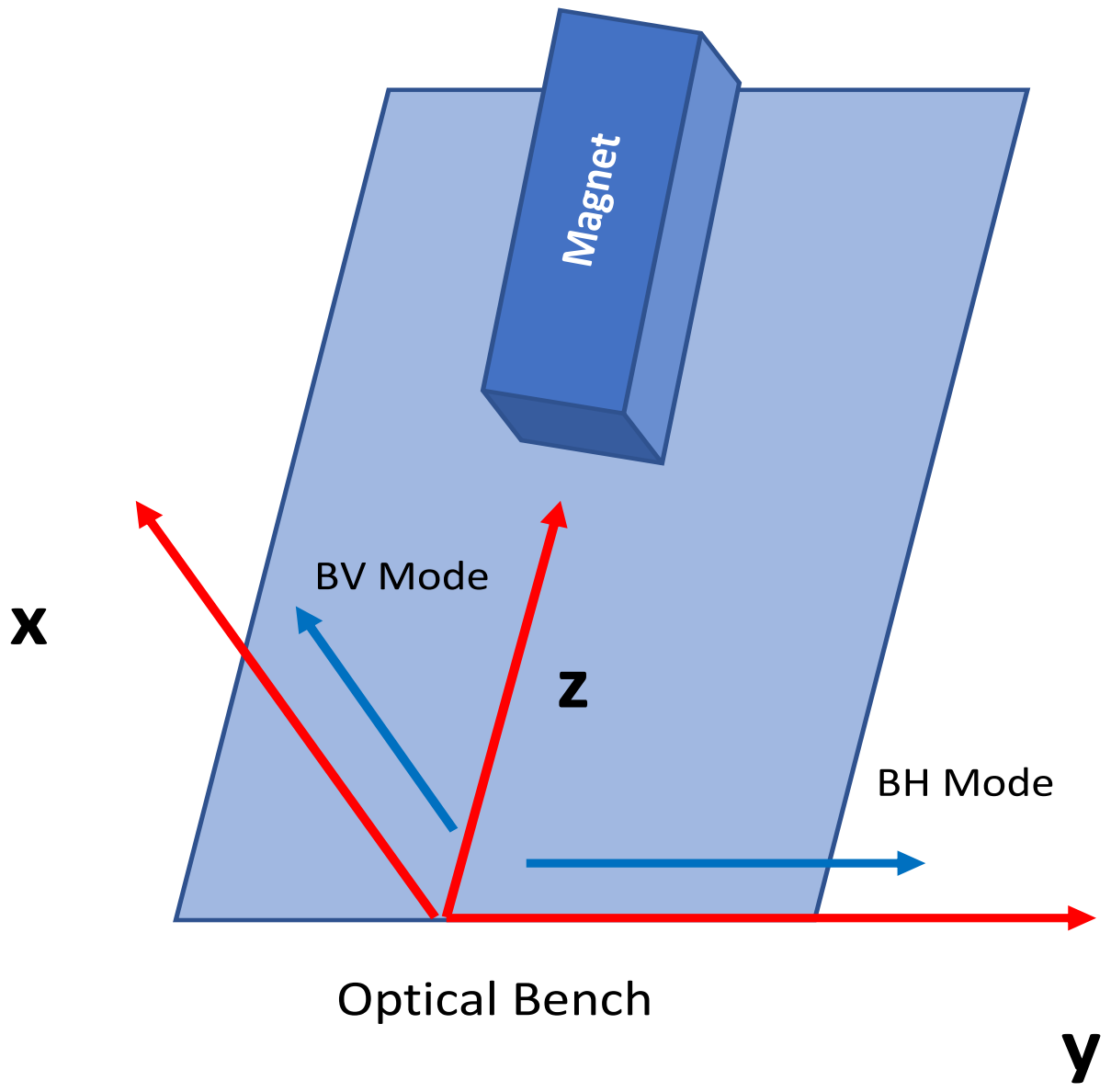

Figure 4: Experimental setup showing the magnetic cavity and direction (z) for beam propagation along with the two orientations for the magnetic field within the cavity (BH and BV) which are referenced throughout this paper.

The laser beam propagates along the z-direction and is polarized along the y-direction. The magnet is rotated such that it is aligned with the laser polarization ("BH") or perpendicular to the laser polarization ("BV").  The magnet is rotated only once to switch modes and is in the BH or BV configuration throughout data taking for any given set of experiments.

Table 1 in the Appendix of this paper gives the experimental parameters for the current setup.  These parameters can be used to calculate a final cross-section for the observed effects. They are also important to reproduction/confirmation of the exact measurements.  As the measurement involves a product of two cross-sections, calculating values for $g_{\alpha\gamma\gamma}$ and $g_{\alpha n}$ requires further measurements, the product of the associated cross sections are in Appendix 3.

**Data Analysis:**

As a function of energy, the decay counts were integrated over a total of 60 hours. During this time period, the system was alternated each hour between the sP and sD modes. This was done to minimize any background differences between the modes. The background was monitored concurrently by a second detector, to set the amplitude for the No Source mode data. The difference between the two modes along with a statistical analysis is shown below.

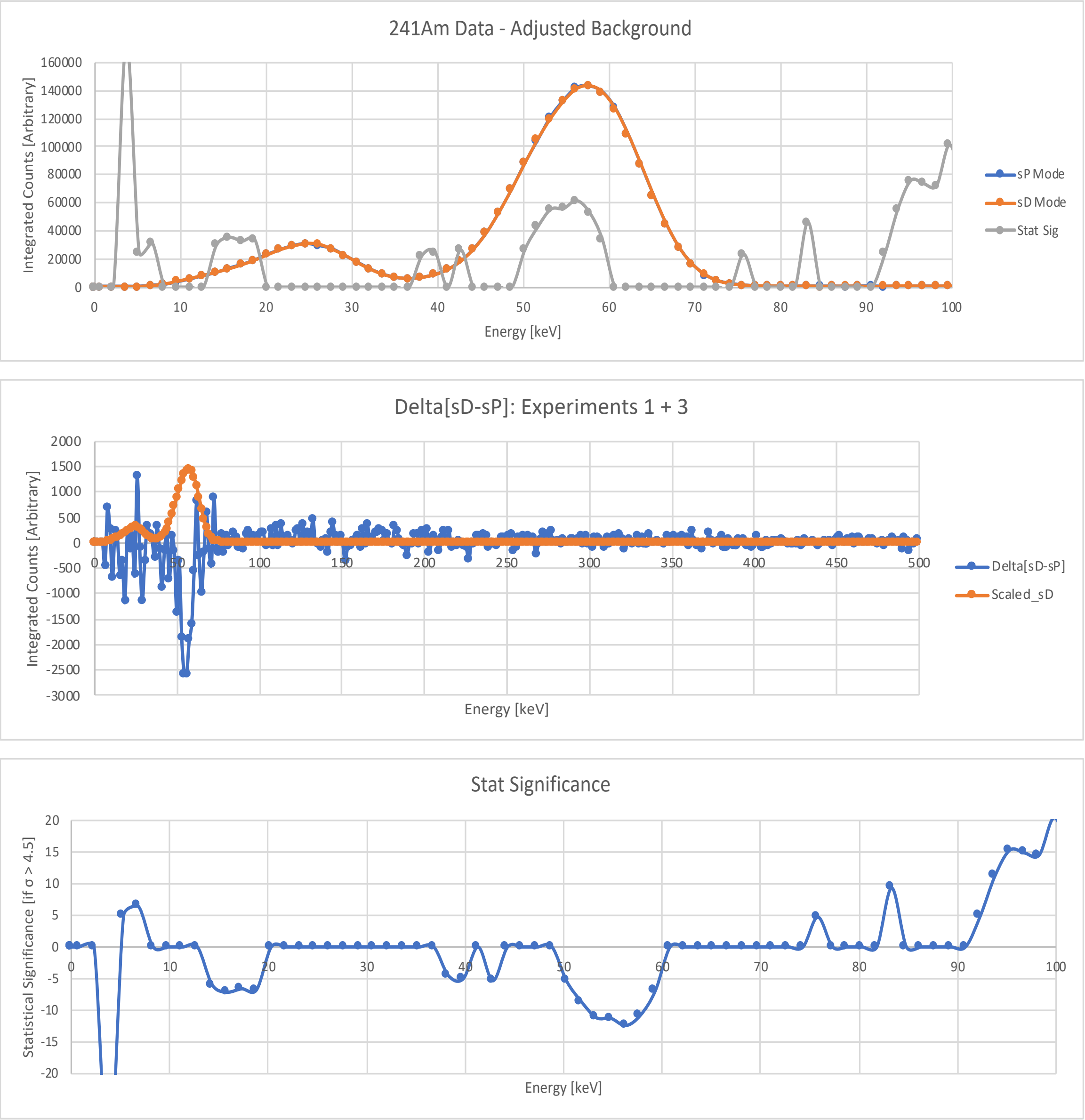

Figure 5: Data showing Integrated Counts over 18hours for Dark (sD) and Light (sP) modes for BH setup with arbitrarily scaled values for the statistical significance overlaid (top panel), sD – sP Raw Counts (middle panel) and the statistical significance, calculated by using a sliding window of width 12-bins equal to width of 59.54 gamma peak, and requiring that $\sigma > 4.5$ (bottom panel).

If the data showed no significant difference between the two modes, if $\sigma < 4.5$, then it was assigned a Statistical Significance value of zero as seen in Figure 5 (bottom panel). The analysis proceeded by establishing first the full-width at half max for the 59.54 keV gamma peak. A sliding window was applied using the same width, 12-bins of the detector, to the difference between the modes. This established that near the 59.54 keV peak the two modes seem to differ significantly as shown. After 18 hours, a 6.49$\sigma$ result was observed at approximately the center of the 59.54 keV peak. This suggests that the $^{241}$Am is responding to some difference between the sP and sD.

The same analysis was repeated for the BV case.

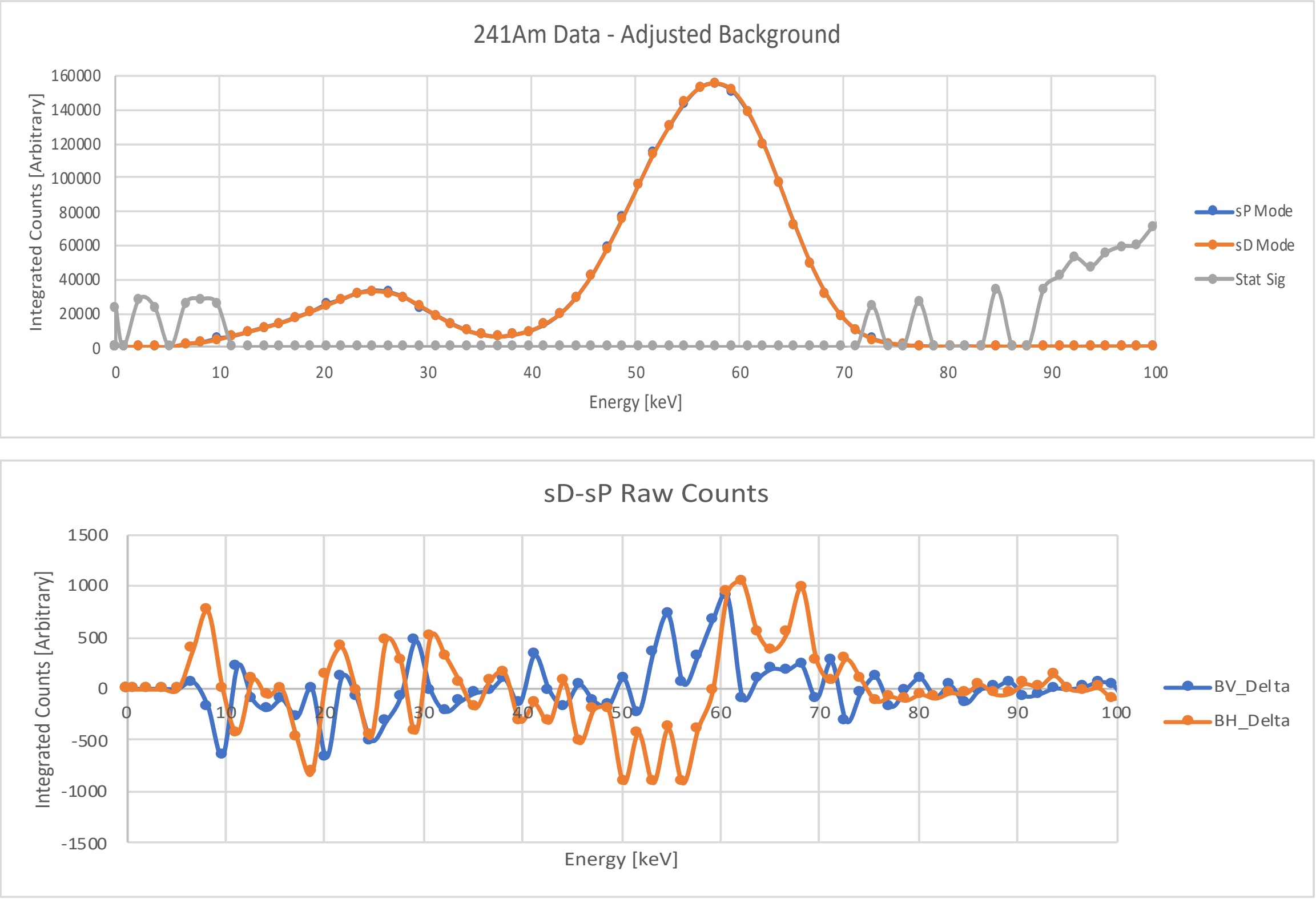

Figure 6: Data showing Integrated Counts over 30hours for each mode including Dark (sD) and Light (sP) and including both BH and BV setups with arbitrarily scaled values for the statistical significance (if $\sigma > 4.5$) overlaid (top panel) and sD – sP Counts with scaled sD (bottom panel).

In theory, if aligning the photon polarization with an external magnetic field creates some weakly interacting beam of scalar particles, then whenever the fields are perpendicular there should be no difference between sP and sD. This means that particles like the axion are their own anti-particle.  If the particle is not its own anti-particle, then the theory would need to consider that both directions of an external field could lead to a non-zero signal.

**Conclusion:**

The current data set shows a statistical significant difference between the Dark (sD) and Light (sP) modes when the photon polarization is aligned with the direction of the external magnetic field (BH).  The other data case, with polarization perpendicular to the B field (BV), showed no statistical significant difference between the sD and sP modes for data taken near the two $^{241}$Am gamma peaks.  However, the differences BH-BV showed sigma values > 8.0. It should also be noted that the BV data was only 18hours while the BH data was over 40hours.

These observations may be the first indications of the production, via the Primakoff mechanism, of weakly interacting particles that influence the decay of $^{241}$Am!

**Appendix 1:**

The parameters for the experiment described in this article are shown below in Table 1. In addition to the $^{241}$Am source, experiments were performed on both $^{238}$U and $^{137}$Cs.  As U and Cs have numerous prominent gamma lines, that overlap due to the detector resolution, the Am source was selected as the focus of this paper.

Table 1: Experimental Parameters for Apparatus Used

| Experimental Parameters 1: | | | | | | | | | | |
|---|---|---|---|---|---|---|---|---|---|---|
| **B Field Max [mT]** | | **Field Length [m]** | | **Lasers** | | **Source Cnts/Sec** | | **Target [mm]** | **Time** | |
| B_1 | 80 | **$\ell$_1** | 1 | Intensity | 1mW (x6) | $^{241}$Am | ~ 100 | 1 | Interval | 1hr |
| B_2 | 50 | **$\ell$_2** | 1 | Wavelength | 632nm | $^{238}$U | ~ 2950 | 4 | Integrated | 60hr |
| B_3 | 50 | **$\ell$_3** | 1 | Cavity | - | $^{137}$Cs | ~ 3150 | unknown | | |

**Appendix 2:**

A second experiment was run using a new setup with a single magnetic field, a blue laser, a mirror cavity in lieu of multiple lasers and a different set of NaI detectors to monitor both the radioactive sample as well as the backgrounds in the room.  The experimental parameters for this second setup are given in Table 2. ***Note:*** the time interval for switching between the sD and sP modes was only 15min each for this second experiment.

Table 2: Experimental Parameters for Second Experiment

| Experimental Parameters 2: | | | | | | | | | | | |
|---|---|---|---|---|---|---|---|---|---|---|---|
| **B Field Max [mT]** | | **Field Length [m]** | | **Lasers** | | **Source Cnts/Sec** | | **Target [mm]** | **Time** | |
| B_1 | 130 | ***ℓ_1*** | 1 | Intensity | 1mW (x1) | $^{241}$Am | ~ 100 | 1 | Interval | 15min |
| B_2 | - | ***ℓ_2*** | - | Wavelength | 445nm | $^{238}$U | - | - | Integrated | 48hr |
| B_3 | - | ***ℓ_3*** | - | Cavity | 2 | $^{137}$Cs | - | - | | |

Only the 241Am source, used in the first setup above, was kept and used in the second experiment. Figure 7 show the results of the second experiment, where the data from the Horizontal and Vertical B field configurations are combined.

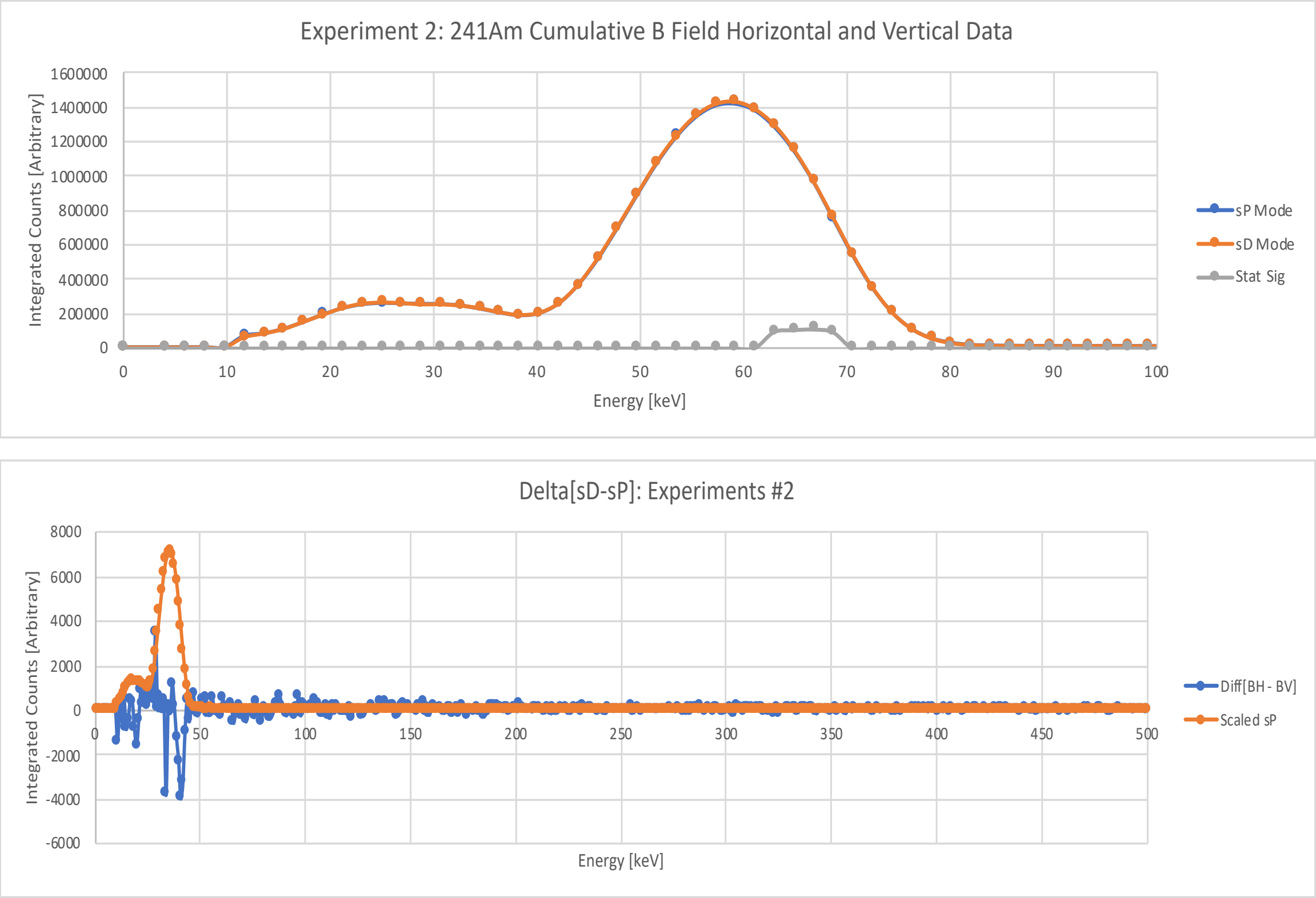

Figure 7: Data showing Integrated Counts over 24hours for each mode including Dark (sD) and Light (sP) and including both BH and BV setups with arbitrarily scaled values for the statistical significance (if $\sigma > 4.5$) overlaid (top panel) and sD – sP Counts with scaled sD (bottom panel).

The statistical significance calculation involved using the same approach as with the first experiment, namely application of a sliding window to the sum over the difference, ( sD – sP ). The window size was established by using the number of bins consistent with a full width at half max for the 59.24keV peak.  Figure 7 shows that there were several points with Sigma > 4.5. At an Energy of 66.84keV, to the right of the center of the 59.54keV peak, sigma reached 5.49.

**Appendix 3:**

Define Delta[Decay] as the difference in decay counts for the value of (sD – sP), for either $^{241}$Am peak.  Then Equation 1.1 can be used to calculate the combined cross section, $\sigma_{\alpha\gamma\gamma}$ x $\sigma_{\alpha n}$, from the measured values of Delta[Decay]:

$$Delta[Decay] = I_{laser} \cdot time \cdot B \cdot \sigma_{\alpha\gamma\gamma} \cdot \sigma_{\alpha n} \cdot t \cdot \ell \cdot \frac{N_A}{m_{241Am}} \cdot f_{geom} \qquad 1.1$$

Where $I_{laser}$ is the number of photons/sec, $B$ gives the magnetic field strength in T, $\sigma_{\alpha\gamma\gamma}$ is the cross section for the creation of a weakly interacting particle, $\sigma_{\alpha n}$ is the cross section for the absorption of the particle by a nucleus, $t$ is the thickness of the target material, $N_A$ is Avagadro's number and $m_{241Am}$ is the gram molecular weight for $^{241}$Am, $\ell$ is the length of the B Field and $f_{geom}$ is a geometric factor to account for the ultra-relativistic production of weakly interacting particles with $m_\alpha \sim 10^{-3} - 10^{-6} eV$ produced by a photon with energy ~ 1eV – currently this factor is taken as 1.  Equation 1.2 gives the estimate for $\sigma_{\alpha\gamma\gamma} \cdot \sigma_{\alpha n}$ based the data:

$$\sigma_{\alpha\gamma\gamma} \cdot \sigma_{\alpha n} = \frac{Delta[Decay]}{I_{laser} \cdot time \cdot B \cdot t \cdot \ell} \cdot \frac{m_{241Am}}{N_A} \cdot \frac{1}{f_{geom}} \qquad 1.2$$

Table 3 gives estimates for the extracted cross sections based on the measurement of the 59.54keV decay peak ONLY. Note: Experiment 2 was integrated over a time that is x2.5 longer than for Experiment 1, but the other experimental parameters, B Field strength and Laser Intensity, gave a reduction of ~ 4.0.

Table 3: Combined Cross Section from Data

| $\sigma_{\alpha\gamma\gamma} \cdot \sigma_{\alpha n}$ **From Time Integrated Experiments:** | | | |
|---|---|---|---|
| | **Experiment 1: BV** | **Experiment 1: BH** | **Experiment 2: BH - BV** |
| **Delta[Decay] ~** | -550 | -15167 | -6300 |
| $\boldsymbol{\sigma_{\alpha\gamma\gamma} \cdot \sigma_{\alpha n}}$ | - | $5.95\ x\ 10^{-38} \pm 0.28\ x\ 10^{-38}$cm$^2$ | $4.52\ x\ 10^{-38} \pm 0.75\ x\ 10^{-38}$cm$^2$ |